\documentclass[preprint,showpacs,superscriptaddress,preprintnumbers,amsmath]{revtex4}

\usepackage{graphicx}
\usepackage{dcolumn} % Align table columns on decimal point
\usepackage{amsfonts}
\usepackage{amssymb}
\usepackage{bm} % bold math
\usepackage[figuresright]{rotating}

\newcommand{\pd}{(p,d)}
\makeatletter
\def\leq{\mathrel{\mathpalette\gl@align<}}
\def\geq{\mathrel{\mathpalette\gl@align>}}
\def\gl@align#1#2{\lower.6ex\vbox{\baselineskip\z@skip\lineskip\z@
    \ialign{$\m@th#1\hfil##\hfil$\crcr#2\crcr\sim\crcr}}}
\makeatother

\begin{document}

%\preprint{ver 7}

\title{Possible evidence of tensor interactions in $^{16}$O observed 
  via \pd~reaction}

\author{H.~J.~Ong}
\email{onghjin@rcnp.osaka-u.ac.jp}
\affiliation{Research Center for Nuclear Physics, Osaka University, Ibaraki, 
              Osaka 567-0047, Japan}
\author{I.~Tanihata}
\affiliation{Research Center for Nuclear Physics, Osaka University, Ibaraki, 
              Osaka 567-0047, Japan}
\affiliation{School of Physics and Nuclear Energy Engineering, 
              Beihang University, Beijing, 100191, China}
\author{A.~Tamii}
\affiliation{Research Center for Nuclear Physics, Osaka University, Ibaraki, 
              Osaka 567-0047, Japan}
\author{T.~Myo}
\affiliation{Research Center for Nuclear Physics, Osaka University, Ibaraki, 
              Osaka 567-0047, Japan}
\affiliation{General Education, Faculty of Engineering, 
              Osaka Institute of Technology, Osaka 535-8585, Japan}
\author{K.~Ogata}
\affiliation{Research Center for Nuclear Physics, Osaka University, Ibaraki, 
              Osaka 567-0047, Japan}
\author{M.~Fukuda}
\affiliation{Department of Physics, Osaka University, Toyonaka, 
              Osaka 560-0043, Japan}
\author{K.~Hirota}
\affiliation{Research Center for Nuclear Physics, Osaka University, Ibaraki, 
              Osaka 567-0047, Japan}
\author{K.~Ikeda}
\affiliation{RIKEN Nishina Center, Wako, 
              Saitama 351-0198, Japan}
\author{D.~Ishikawa}
\affiliation{Research Center for Nuclear Physics, Osaka University, Ibaraki, 
              Osaka 567-0047, Japan}
\author{T.~Kawabata}
\affiliation{Department of Physics, Kyoto University, 
              Kitashirakawa, Kyoto 606-8502, Japan}
\author{H.~Matsubara}
%\thanks{Present address: RIKEN Nishina Center, Wako, 
%              Saitama 351-0198, Japan}
\affiliation{Research Center for Nuclear Physics, Osaka University, Ibaraki, 
              Osaka 567-0047, Japan}
\author{K.~Matsuta}
\affiliation{Department of Physics, Osaka University, Toyonaka, 
              Osaka 560-0043, Japan}
\author{M.~Mihara}
\affiliation{Department of Physics, Osaka University, Toyonaka, 
              Osaka 560-0043, Japan}
\author{T.~Naito}
\affiliation{Research Center for Nuclear Physics, Osaka University, Ibaraki, 
              Osaka 567-0047, Japan}
\author{D.~Nishimura}
%\thanks{Present address: Department of Physics, Tokyo University of Science, 
%  Noda, Chiba 278-8510, Japan}
\affiliation{Department of Physics, Osaka University, Toyonaka, 
              Osaka 560-0043, Japan}
\author{Y.~Ogawa}
\affiliation{Research Center for Nuclear Physics, Osaka University, Ibaraki, 
              Osaka 567-0047, Japan}
\author{H.~Okamura}
\affiliation{Research Center for Nuclear Physics, Osaka University, Ibaraki, 
              Osaka 567-0047, Japan}
\author{A.~Ozawa}
\affiliation{Institute of Physics, Tsukuba University, Tsukuba,
              Ibaraki 305-8571, Japan}
\author{D.~Y.~Pang}
\affiliation{School of Physics and Nuclear Energy Engineering, 
              Beihang University, Beijing, 100191, China}
\author{H.~Sakaguchi}
\affiliation{Research Center for Nuclear Physics, Osaka University, Ibaraki, 
              Osaka 567-0047, Japan}
\author{K.~Sekiguchi}
\affiliation{Department of Physics, Tohoku University, Sendai, 
              Miyagi 980-8578, Japan}
\author{T.~Suzuki}
\affiliation{Research Center for Nuclear Physics, Osaka University, Ibaraki, 
              Osaka 567-0047, Japan}
\author{M.~Taniguchi}
\affiliation{Department of Physics, Nara Women's University, 
              Nara 630-8506, Japan}
\author{M.~Takashina}
\affiliation{Research Center for Nuclear Physics, Osaka University, Ibaraki, 
              Osaka 567-0047, Japan}
\author{H.~Toki}
\affiliation{Research Center for Nuclear Physics, Osaka University, Ibaraki, 
              Osaka 567-0047, Japan}
\author{Y.~Yasuda}
\affiliation{Research Center for Nuclear Physics, Osaka University, Ibaraki, 
              Osaka 567-0047, Japan}
\author{M.~Yosoi}
\affiliation{Research Center for Nuclear Physics, Osaka University, Ibaraki, 
              Osaka 567-0047, Japan}
\author{J.~Zenihiro}
%\thanks{Present address: RIKEN Nishina Center, Wako, 
%              Saitama 351-0198, Japan}
\affiliation{Research Center for Nuclear Physics, Osaka University, Ibaraki, 
              Osaka 567-0047, Japan}
%\collaboration{RCNP-E314 Collaboration}

\date{\today}% It is always \today, today,

\begin{abstract}
%%0123456789012345678901234567890123456789012345678901234567890123456789
  We have measured $^{16}$O\pd~reaction using 198-, 295- and 392-MeV proton 
  beams to search for a direct evidence on the effect of the 
  tensor interactions in light nucleus.  Differential cross sections of the 
  one-neutron transfer reactions populating the ground states and several 
  low-lying excited states in $^{15}$O were measured.  Comparing the ratios 
  of the cross sections for each excited state to the one for the ground state 
  over a wide range of momentum transfer, we found a marked enhancement for 
  the positive-parity state(s).  The observation indicates large components
  of high-momentum neutrons in the initial ground-state configurations, due 
  possibly to the tensor interactions.
\end{abstract}

     % 21.30.-x  Nuclear forces
     % 24.50.+g  Direct reactions
     % 25.40.Hs  Nucleon-induced transfer reaction
\maketitle

  Tensor interactions are some of the most important nuclear interactions 
  acting between two nucleons.  The tensor interactions, 
  originate mainly from the pion-exchange interactions, provide the most 
  significant attraction in nuclear interactions.  The necessity to include 
  tensor interactions in theoretical calculations to reproduce the quadrupole 
  moment~\cite{schwinger-kellog} as well as the binding 
  energy~\cite{bethe-rarita} of the deuteron, which is the simplest and only 
  stable two-nucleon composite system, affords decisive evidences on the 
  importance of the tensor interactions.  The tensor interactions provide 
  $70\sim 80\%$ of the attractive interactions~\cite{ericson,ikeda} in 
  deuteron and induce nucleons with high momenta~\cite{ikeda} through the 
  D-wave component.

  Besides the deuteron, earlier theoretical studies~\cite{gerjuoy-schwinger} 
  had also pointed out the importance of the tensor interactions to the binding
  of three- and four-nucleon systems, accounting for almost 50$\%$ of the 
  nuclear attraction.  Several experiments using polarized deuteron beams had
  since been performed to measure the tensor analyzing powers for stripping 
  reactions~\cite{knutson-roman-karp} and deuteron capture 
  reaction~\cite{weller}, providing evidences on the existence of the D-state
  components in the $^3$H and $^{3,4}$He.

  For heavier nuclei, recent ab-initio calculations~\cite{ab-initio} on light 
  nuclei also show essential importance of the tensor interactions for binding 
  nuclei up to mass number $A=12$.  The pion exchange interactions, in which 
  the tensor interactions are the dominant components, constitute 70--80$\%$ of 
  the whole two-body potentials.  In addition, detailed 
  studies on experimental data~\cite{ozawa} and the subsequent theoretical 
  studies~\cite{otsuka} have indicated a possible important role of the tensor
  interactions in changing the magic numbers and the orders of single-particle 
  orbitals in neutron-rich nuclei, although the strength of the tensor 
  interactions in the shell-model space is not large, and is treated only as 
  a perturbation.  More recently, theoretical calculations on $^{9-11}$Li that 
  include explicitly the tensor interactions have pointed out~\cite{myo-lithium}
  the importance of the tensor interactions in understanding the structure of 
  those nuclei, and predicted high momentum components in the ground states.  
  The results offered a possible intriguing explanation to the development of 
  the neutron-halo structure through the Pauli blocking effect in $^{11}$Li.

  Experiments using the electron~\cite{eeprime,leuschner} or 
  proton-induced~\cite{p2p} knockout reaction had been performed to probe 
  the tensor correlations in nuclei from $^{12}$C to $^{208}$Pb.  However, 
  since it is difficult to isolate the tensor effects unambiguously in 
  these experiments due to the presence of other correlations such as 
  the short-range repulsion, alternative methods that could provide more 
  direct experimental evidences are called for.

  In this paper, we report a possible direct observation of the tensor-force 
  effect in the ``doubly-closed-shell'' $^{16}$O using the one-neutron transfer
  \pd~reaction.  The tensor interactions mix large orbital angular momentum 
  states, giving rise to high momentum components through D-wave component in 
  the relative coordinate of two nucleons in finite nuclei.  In fact, recent 
  theoretical calculations~\cite{neff,horiuchi,schiavilla} have predicted 
  enhanced momentum distributions at around 2 fm$^{-1}$ due to the tensor 
  interactions.  In this work, we measured the cross sections of the 
  one-neutron pickup reaction at momentum transfer around 2 fm$^{-1}$ by 
  observing the ground state as well as excited states in $^{15}$O.  We found 
  strong relative enhancements of the cross section to the positive-parity 
  excited state(s) (the $\frac{1}{2}^+$ and/or $\frac{5}{2}^+$ states)
  at high momentum transfer.

  In the independent single particle model (shell model), the ground state of 
  $^{16}$O consists of eight protons and eight neutrons filling up the 
  1s$_{1/2}$,  1p$_{3/2}$ and 1p$_{1/2}$ orbitals.  Hence, the positive-parity 
  states in $^{15}$O can be reached through direct neutron-pickup reaction
  only if the $^{16}$O ground state has an admixture of 1d$_{5/2}$ or 2s$_{1/2}$ 
  state, in the absence of multi-step processes.  Such admixture is possible if
  the tensor interactions play a dominant role, since the tensor interactions 
  induce changes in the total orbital and spin angular momenta by 
  $|\Delta L|$=2 and $|\Delta S|$=2, giving rise to two-particle two-hole 
  ($2p2h$) configurations which include (1p$_{3/2}$)$^{-2}$(1d$_{5/2}$)$^2$ and 
  (1p$_{1/2}$)$^{-2}$(2s$_{1/2}$)$^2$ in the ground state of 
  $^{16}$O~\cite{myo-ogawa}.

  The \pd~reaction has been applied extensively to study the single particle 
  nature of nuclei.  In this reaction, a neutron is picked up from the target 
  nuclei to form a deuteron.  The advantage of this reaction lies in the 
  selectivity of the momentum of the picked-up neutron.  Under the single-step
  pickup reaction using a deuteron target, the momentum of the picked-up 
  neutron in the target deuteron is equivalent to the momentum transfer, 
  namely the difference between the momenta of the outgoing deuteron and 
  the incident proton $\vec{P}_{\rm d}$-$\vec{P}_{\rm p}$.  Neutron pickup 
  reactions with a nuclear target, when a deuteron is observed at small 
  scattering angles, are expected to occur under the same reaction mechanism 
  and thus can be used to extract spectroscopic information on the neutron 
  residing in the target nucleus.

% Experiment  
  The experiment was performed at the WS beamline of the RCNP cyclotron
  facility.  Proton beams at $E_p$ = 198, 295 and 392 MeV were provided by 
  the RCNP ring cyclotron and transported in the achromatic mode 
  to a target placed in a scattering chamber.  The typical beam spot size 
  at the target was 1 mm in diameter.

  We used a windowless and self-supporting thin ice sheet~\cite{kawabata} as 
  the target.  The thin ice sheet, which was made of pure water, was cooled by 
  liquid nitrogen and kept below 140 K throughout the experiment.  
  A new ice target was prepared before each measurement with different proton 
  beam energy to reduce $^{12}$C contaminants from vacuum pump oil.  The 
  thicknesses of the targets were determined to be $32\pm 2$, $30\pm 3$ and 
  $62\pm 5$ mg/cm$^2$ for the measurements with $E_p$ = 198, 295 and 392 MeV 
  respectively, through measurement of the elastic scattering off the
  hydrogen.

  The deuterons produced in the one-neutron pickup reactions were momentum 
  analyzed by the Grand Raiden spectrometer~\cite{gr} and detected by two 
  multi-wire drift chambers and two 10-mm thick plastic scintillation 
  detectors placed at the exit focal plane about 20 m from the target.
  The acceptance of the scattered deuterons was limited to 40 mrad horizontally
  and 60 mrad vertically using a collimator slit.  The deuterons were 
  identified using the time-of-flight information and the pulse-height 
  information from the two plastic scintillators.  The momenta were determined 
  based on the horizontal position information obtained with the drift chambers
  and the strength of the magnetic field.

  To cover momentum transfer at around 2 fm$^{-1}$, we have performed
  measurements at several finite angles ($\theta_{\rm lab}$) from 5$^\circ$ to 
  25$^\circ$.  The proton beam exiting the target was stopped in a Faraday 
  cup.  The beam current was monitored throughout the experiment using a 
  current integrator connected directly to the Faraday cup. 

% Results  
  \begin{figure}[htbp]
    \begin{center}
      \includegraphics[scale=0.48]{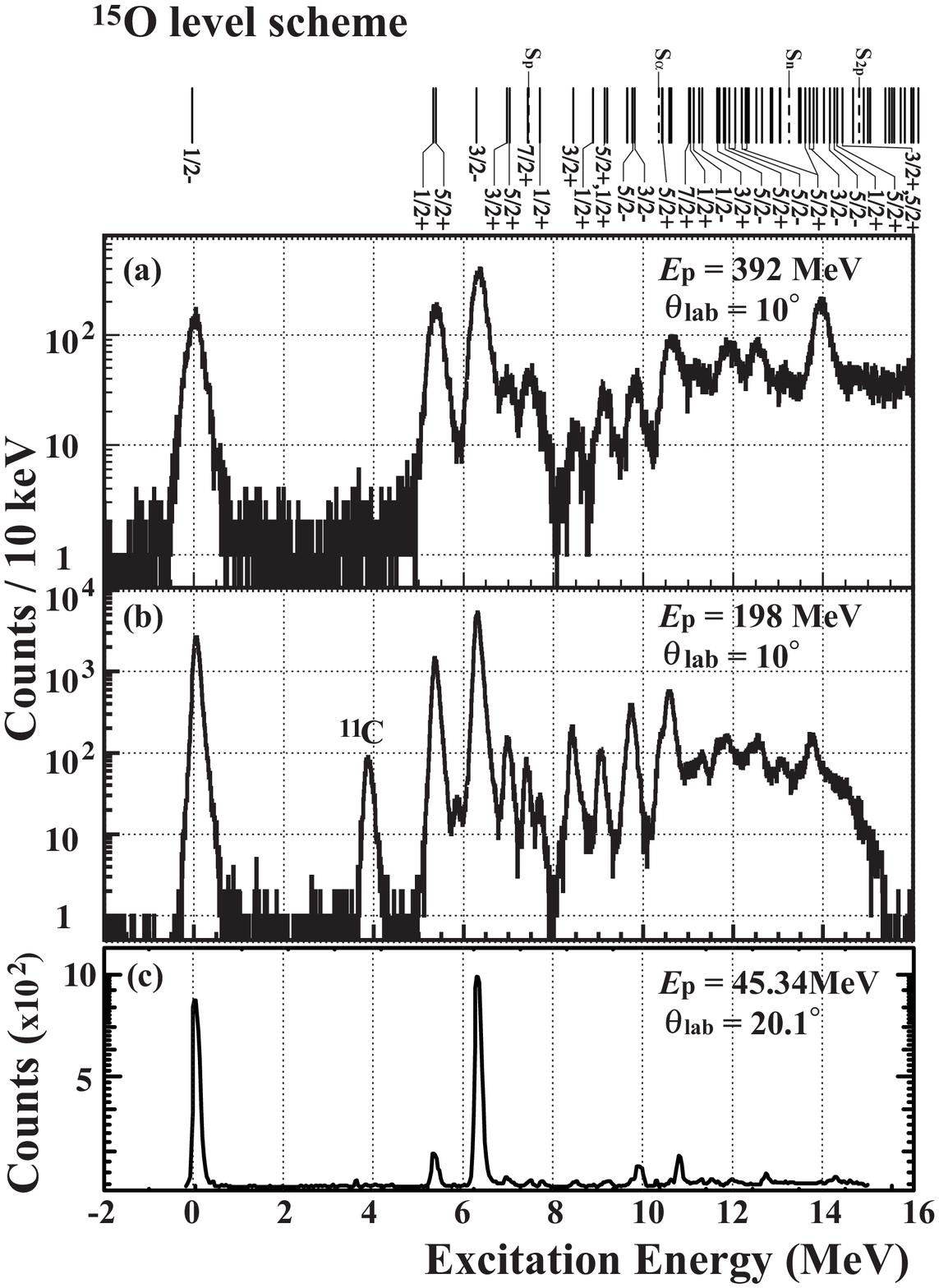}
      \caption{Typical excitation energy spectra for the $^{16}$O(p,d)$^{15}$O 
        reactions obtained at proton energies (a) 392 MeV and (b) 198 MeV.  
        The deuterons were detected at 10$^\circ$ with respect to the incident 
        beam.  The level scheme for $^{15}$O is shown at the top for reference.
        For clarity, only the well-established spin-parities are shown.  
        (c) A spectrum for $E_{\rm p} =$ 45.34 MeV replicated from 
        the figure in ref.~\cite{snelgrove} is shown for comparison.} 
      \label{energy_spectra}
    \end{center}
  \end{figure}

  The excitation energy spectra of the $^{16}$O\pd$^{15}$O reaction were 
  reconstructed using the information of the proton beam energy as well as 
  the scattering angle and the measured momenta of the deuterons.  
  Figure~\ref{energy_spectra}(a) and (b) show the excitation energy spectra 
  for the reactions obtained with proton beams at 392 and 198 MeV, where 
  the deuterons were detected at 10$^\circ$ with respect to the incident beam.
  The measurement time was about an hour and the beam intensity was about 2 nA 
  for each measurement.  Several peaks corresponding to the ground state 
  as well as the excited states of the residual nuclei $^{15}$O were clearly 
  observed.  For reference, the level scheme is shown at the top of the figure.
  The 5.183-MeV, $\frac{1}{2}^+$ and the 5.240-MeV, $\frac{5}{2}^+$ states in 
  $^{15}$O were not resolved in the present experiment.  Nonetheless, 
  this should not alter our conclusion.  For comparison, 
  the energy spectrum for $E_{\rm p} =$ 45.34 MeV at 20.1$^\circ$ replicated 
  from the figure in ref.~\cite{snelgrove} is also shown 
  (Fig.~\ref{energy_spectra}(c)).

  Since the ground $\frac{1}{2}^-$ and the 6.176-MeV excited $\frac{3}{2}^-$ 
  states in $^{15}$O can be assumed to be neutron p$_{1/2}$ and p$_{3/2}$ hole 
  states, one expects such states to be relatively strongly populated through 
  direct pickup of a neutron.  It is, however, surprising that the 
  positive-parity states are also strongly populated in the $^{16}$O\pd$^{15}$O 
  reaction.  In particular, the population of the $\frac{5}{2}^+$ or 
  $\frac{1}{2}^+$ state indicates possible contribution from the sd-shell.  
  At low energy, as shown in Fig.\ref{energy_spectra}(c), the cross section to
  the positive-parity state is smaller than those to the negative-parity 
  states.

  In general, the cross sections of all states diminish with increased proton
  beam energy, i.e. increased momentum transfer, due to momentum mismatching.  
  This trend is particularly pronounced for the ground state, which is an 
  evidence of diminishing high-momentum neutrons in the initial ground-state 
  configurations.  Notice that the relative intensity of the positive-parity 
  states around 5.2 MeV in $^{15}$O increases at very large momentum transfer, 
  which apparently indicates slower decrease or relative increase compared 
  with the negative-parity states.

  \begin{figure}[htbp]
    \begin{center}
      \includegraphics[scale=0.35]{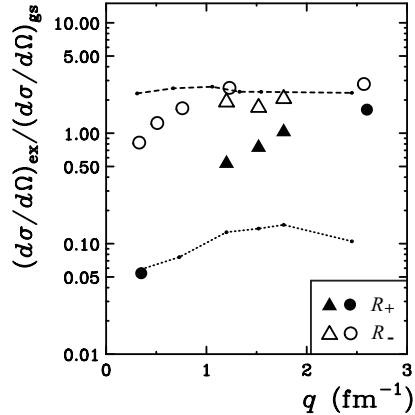}
      \caption{Ratios of the intensities of the $\frac{3}{2}^-$ (open symbols) 
        and the positive-parity ($\frac{5}{2}^+$ and/or $\frac{1}{2}^+$) 
        (filled symbols) excited states to that of the ground $\frac{1}{2}^-$ 
        state as functions of momentum transfer.  The present data are 
        represented by the filled and open triangles.  Other data are taken 
        from previous works with 45-, 65-, 100-, 200- and 800-MeV proton 
        energies.  The dashed (dotted) curve represents the ratios of the 
        1p$_{3/2}$ (1d$_{5/2}$) and 1p$_{1/2}$, obtained by zero-range CDCC-BA 
        calculations with finite-range correction using the Dirac 
        phenomenological potentials.}
      \label{ratio}
    \end{center}
  \end{figure}
  To examine the relative strength of the excited states, we divided the cross 
  sections for the excited states by that of the ground state.  The ratios thus 
  obtained were plotted against the averaged momentum transfer as shown in 
  Fig.~\ref{ratio}.  In order to avoid possible complications due to reaction 
  mechanism, we have confined ourselves to the data at 10$^\circ$.  
  The filled symbols represent the ratios (denoted by $R_+$) for the 
  positive-parity ($\frac{5}{2}^+$ and/or $\frac{1}{2}^+$) states, while the 
  open symbols represent the ratios (denoted by $R_-$) for the negative-parity,
  $\frac{3}{2}^-$ state.  The filled and open triangles are the data obtained 
  in the present work.  The error bars, which mainly consist of the statistical 
  errors ($<3\%$), are smaller than the symbols.  Note that the contributions 
  from the $^{12}$C contaminant have been estimated and subtracted.  The 
  acceptance and detection efficiency of the deuterons corresponding to 
  the ground and excited states were almost constant.  Other data were taken 
  from the previous measurements at proton energies of 
  45.34 MeV~\cite{snelgrove}, 65 MeV~\cite{roos}, 100 MeV~\cite{lee}, 
  200 MeV~\cite{aberg} and 800 MeV~\cite{smith}.  
  As evidence from the figure, the ratios for the positive-parity states 
  increase drastically by a factor of 30 from $q$ transfer 0.3 fm$^{-1}$ 
  to 2.6 fm$^{-1}$, whereas the ones for the negative-parity states only triple
  over the same momentum-transfer range.

% Discussion
  The energy dependence of the differential cross sections at the first $l$=1 
  maxima, which lie between 10$^\circ$ and 20$^\circ$ in the center-of-mass 
  frame close to 10$^\circ$ at the laboratory frame, has been reported for 
  proton energies from 18.5 MeV to 100 MeV~\cite{snelgrove}.  
  The present data, together with the data at $E_p$=800 MeV, indicate that 
  $R_-$ remains almost constant ($\sim 3.7/1.68$, see below) above 
  $q$=1 fm$^{-1}$.

  To investigate the relative strengths, we performed theoretical calculations 
  to obtain the ratios for the $\frac{3}{2}^-$ and $\frac{1}{2}^-$ states at 
  10$^\circ$.  The dashed curve in Fig.~\ref{ratio} show the calculated ratios 
  for proton energies from 45 to 800 MeV, obtained with the 
  Continuum-Discretized Coupled-Channel method with Born Approximation to 
  the transition operator $\hat{V}_{\rm tr}$ for the transfer, i.e. 
  CDCC-BA~\cite{cdcc} calculation.  We made zero-range approximation with
  finite-range correction to $\hat{V}_{\rm tr}$, and used nucleon-nucleus 
  distorting potentials based on the Dirac phenomenology~\cite{dirac-pheno}.  
  In obtaining the calculated ratios, we have adopted 
  the shell-model spectroscopic factors of 1.68 and 3.7~\cite{alexbrown} for 
  the p$_{1/2}$ and p$_{3/2}$ states, respectively.  The calculations are 
  qualitatively consistent with the experimental data above $E_p$=100 MeV 
  (or $q \geq$0.8 fm$^{-1}$), indicating that the ratios of the 1p$_{3/2}$ and 
  1p$_{1/2}$ states can be understood within the present shell-model framework. 
  Although not shown in Fig.~\ref{ratio}, calculations using the Adiabatic 
  Distorted-Wave Born Approximation~\cite{johnson-soper} with relativistic 
  correction for the reaction with 200-MeV incident protons also give a ratio 
  consistent with the experimental data.

  For the positive-parity states, calculations using the shell model which 
  include two-particle two-hole (2$p$2$h$) configuration were reported to 
  reproduce the experimental data at proton energy below 45.34 MeV.  The sum 
  spectroscopic factors were found to be as small as 0.15 and 0.02 for the 
  1d$_{5/2}$ and the 2s$_{1/2}$ states respectively~\cite{snelgrove}.  
  Assuming only the 1d$_{5/2}$ orbital and the spectroscopic factor of 0.15, 
  we performed calculations for the positive-parity state.  The ratios,
  which are represented by the dotted curve in Fig.~\ref{ratio}, are almost 
  constant from $q =$1.0 fm$^{-1}$ onwards.  It is obvious that the calculations
  underestimate the experimental data at large momentum transfer by about an 
  order of magnitude.  This result is expected since it is well 
  known~\cite{wall} that the conventional shell model does not supply enough 
  high-momentum components.  Although the analysis of the $^{16}$O(e,e'p)
  reaction data has indicated a 15$\%$ contribution from the multi-step 
  processes~\cite{leuschner} to the spectroscopic factor of the 1d$_{5/2}$ 
  state, it is not sufficient to account for the observed enhancement at 
  large momentum transfer.

  Recently, theoretical calculations using the Tensor-Optimized Shell Model 
  incorporating the Unitary Correlation Operator Method (TOSCOM) that include 
  2$p$2$h$ configurations generated by tensor interactions in the ground state 
  have been performed for $^{4}$He~\cite{myo-alpha} and 
  $^{9,10,11}$Li~\cite{myo-lithium}.  
  The tensor interactions mix the high-momentum components through 
  $|\Delta L|=2$, $|\Delta S|=2$ admixture.  To confirm whether or not the
  observed enhancement of the positive-parity 
  states is due to the effect of the tensor interactions, reaction analysis 
  using the wave functions of $^{16}$O obtained with TOSCOM as well as more 
  experimental data using one-nucleon transfer reactions on other nuclei 
  are anticipated.

  We shall note that the scattering angle of deuteron was set to
  $\ge 10^\circ$ to obtain momentum transfer $\sim$2 fm$^{-1}$, due to the 
  limitation of the proton-beam energy at RCNP.  It would be more desirable 
  to use higher energy proton beam and measure the cross section near 
  0$^\circ$ to minimize possible complications due to reaction mechanisms.

% Summary
  In summary, we have performed an experiment at RCNP, Osaka University using
  the $^{16}$O\pd~reactions with proton beams at 198 MeV, 295 MeV and 392 MeV 
  to search for a direct evidence on the effect of the tensor interactions 
  in light nuclei.  Differential cross sections of the one-neutron transfer 
  reactions populating the ground states as well as several low-lying excited 
  states in $^{15}$O were measured.  By considering the ratio of the cross 
  section for each excited state and the one for the ground state over a wide 
  range of momentum transfer, we have observed a marked enhancement in 
  the ratio for the $\frac{1}{2}^+$ and/or $\frac{5}{2}^+$ state(s) 
  in $^{15}$O.  The result indicates relative increase of high-momentum 
  neutrons in the initial ground-state configuration with neutron(s) in 
  1s$_{1/2}$ and/or 1d$_{5/2}$ orbital, and thus may be a direct signature of 
  the effect of the tensor interactions in $^{16}$O.  The present work shows 
  that one-nucleon transfer reactions, e.g. the \pd~reaction, afford useful 
  means to probe the effect of the tensor interactions in nuclei.

\section*{Acknowledgment}
  We thank the RCNP Ring Cyclotron staff for the stable proton beams throughout 
  the experiment.  H.~J.~O and I.~T. would like to acknowledge the support 
  of Prof. Akihiro Tohsaki (Suzuki) and his spouse which helped to kick-start 
  this project.  This work was supported in part by Grant-in-Aid for 
  Scientific Research No. 20244030, 20740163 and 23224008 from Monbukagakusho, 
  Japan.

\end{document}